\documentclass[twocolumn,showpacs,preprintnumbers,superscriptaddress,amsmath,amssymb]{revtex4}

\usepackage{graphicx}
\usepackage{dcolumn}
\usepackage{bm}
\usepackage{amsmath}
\usepackage{multirow}
\newcommand{\astra}{{\sc Astra }}

\newcommand{\mean}[1]{\mbox{$\langle{#1}\rangle$}}

\begin{document}

\title{Photoinjector-generation of a flat electron beam \\ with
transverse emittance ratio of $100$ }
\author{P. Piot} \email[Electronic address: ]{piot@fnal.gov}
\affiliation{Northern Illinois University,
DeKalb IL 60115, USA} \affiliation{Fermi National Accelerator Laboratory,
Batavia, IL 60510, USA}
\author{Y.-E Sun}
\email[Electronic address: ]{yinesun@uchicago.edu} \thanks{now at
Argonne National Laboratory.} \affiliation{University of Chicago,
Chicago, IL 60637, USA}
\author{K.-J. Kim}
\affiliation{University of Chicago,
Chicago, IL 60637, USA}
\affiliation{Advanced Photon Source, Argonne National Laboratory, Argonne, IL 60439, USA} 
\date{\today}

\begin{abstract}
The generation of a flat electron beam directly from a
photoinjector is an attractive alternative to the electron damping
ring as envisioned for linear colliders. It also has potential
applications to light sources such as the generation of
ultra-short x-ray pulses or Smith-Purcell free electron lasers. In
this Letter, we report on the experimental generation of a
flat-beam with a measured transverse emittance ratio of $100\pm
20.2$ for a bunch charge of $\sim 0.5$~nC; the smaller measured
normalized root-mean-square emittance is $\sim
0.4$~$\mu$m and is limited by the resolution of our experimental
setup. The experimental data, obtained at the Fermilab/NICADD
Photoinjector Laboratory, are compared with numerical simulations
and the expected scaling laws.
\end{abstract}
\pacs{ 29.27.-a, 41.85.-p,  41.75.Fr}
\maketitle
%
%
Flat electron beams, e.g. beams with large transverse emittance ratios, 
have been proposed in the context of linear
colliders and some novel electron-beam-based light sources. In the
case of a linear e$^+$/e$^-$ collider, a flat beam at the interaction
point reduces the luminosity disruption caused by beamsstrahlung~\cite{yokoya}. 
In the case of light sources, such as the LUX project proposed at LBL~\cite{lux}, 
a flat beam with a smaller emittance of 0.3~$\mu$m and emittance ratio of 50 
is needed to produce x-ray pulses that can be compressed to the order of 
femtoseconds via standard x-ray pulse compression techniques~\cite{zholents}. 
Another type of light source recently drawing attention is based on
self-amplification of Smith-Purcell radiation~\cite{smithpurcell}.
Given one or two planar metal gratings, a flat beam could enhance the
interaction between the electrons and metal grating surface, thus reducing 
the gain length associated with the Smith-Purcell free-electron-laser
mechanism~\cite{kjk2,sp,YHzhang}.

In the proposed International Linear Collider (ILC) the needed flat-beam 
parameters (emittance ratio of 300) are foreseen to be achieved via radiation 
cooling in a damping ring~\cite{ilc}. Although the required
transverse emittances for the ILC have been demonstrated at 
the ATF damping ring of KEK~\cite{kek}, ILC puts stringent requirements on 
the damping ring design, and the cost of the damping ring is a
significant portion of the total collider cost. Therefore
alternative ways of producing flat beams directly from an electron
source have been explored by several groups~\cite{rosenzweig}. In
conjunction with the invention of a linear transformation capable
of transforming an incoming flat beam into an
angular-momentum-dominated (or ``magnetized") beam~\cite{derbenev}, a scheme 
which inverses this transformation was proposed to generate a flat beam
directly out of a photoinjector~\cite{brinkmann2}. The method
consists of generating an magnetized beam by immersing the photocathode in an
axial magnetic field. After acceleration, the beam is transformed
into a flat beam using three skew quadrupoles~\cite{BurovDanilov}.  
This has been verified experimentally~\cite{Edwards,Edwardspac01,flat2, yineprstab}, 
and transverse emittance ratios of 40-50 were reported. Theoretical 
analysis of the conversion of a magnetized cylindrically-symmetric beam into
a flat beam has been presented~\cite{BND-PRE,kjk} and
some of the associated limitations explored~\cite{yinesunlinac2004, 
yinesunpac2005}. In the present Letter we report on an improvement
of the experimental conditions and methods that led to a measured
transverse emittance ratio of approximately 100.
\begin{figure*}[htb!!!!!!!!!!!!!!!!!!!!!!!!!]\centering
\includegraphics[width=0.90\textwidth]{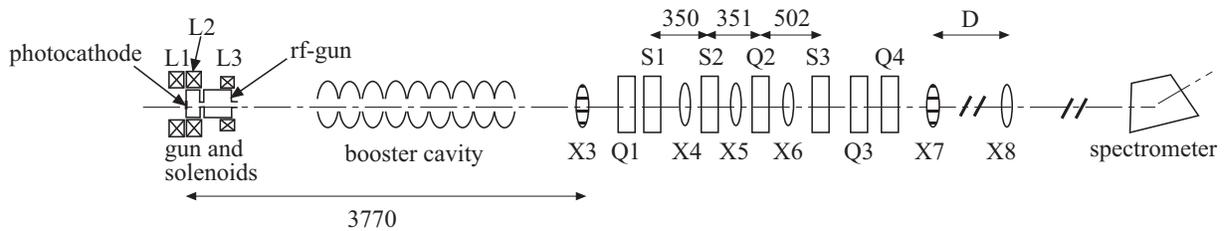}
\caption{Overview of the Fermilab/NICADD photoinjector. ``X" refer to
diagnostics stations (beam viewers, and/or slit location), ``L" to the
solenoidal lenses, ``Q" to quadrupoles and ``S" to the skew quadrupoles.
All distances are in mm, with $D$=800 (or 1850 for the data
presented in Fig.~\ref{fig:emitvsL}).} \label{fig:overview}
\end{figure*}
%
%

The flat-beam experiment was carried out at the
Fermilab/NICADD~\footnote{NICADD is an acronym for Northern
Illinois Center for Accelerator and Detector Development.}
Photoinjector Laboratory (FNPL)~\cite{fnpl}; see
Fig.~\ref{fig:overview} for the layout. In brief, electron bunches
with variable charge ($Q\le$20~nC) are generated via photoemission
from a cesium telluride photocathode located at the back plate of
a 1+1/2 cell radio-frequency (rf) cavity operating at 1.3~GHz (the
``rf gun"). The beam is then accelerated in a 1.3~GHz superconducting
rf cavity (the booster cavity)~\cite{sccavity} to approximately
16~MeV. The rf gun is surrounded by three solenoidal lenses that
are designed to control the beam transverse emittance. For 
flat-beam production the first solenoidal lens (L1) is turned off,
and the two others (L2 and L3) are tuned to provide the desired
magnetic field on the photocathode along with the proper focusing.
The beam is thereby produced in the presence of a significant
axial magnetic field and has an average angular momentum 
given by $\mean{L}=eB_0\sigma_c^2$, where $e$ is the electron
charge, $B_0$ the axial magnetic field on the photocathode
surface, and $\sigma_c$ the root-mean-square (rms) transverse size
of the drive-laser spot on the photocathode.
The transformation of the magnetized beam into a flat beam occurs
downstream of the booster cavity. Three skew quadrupoles (S1, S2,
and S3 in Fig.~\ref{fig:overview}) provide a net torque on the
beam thereby removing its initial angular
momentum~\cite{KJKtorque,yinephD}. The skew quadrupoles are 
henceforth referred to as the ``transformer". Given the incoming $4\times 4$ beam
covariance matrix $\Sigma_0$, the quadrupole strengths are set to
provide the proper transport matrix $M$ so that the covariance
matrix at the exit of the transformer, $\Sigma=M \Sigma_0
\widetilde{M}$ (where the upper tilde denote the transpose),
is block-diagonal. An analytical solution for the quadrupole
settings was derived under the thin-lens approximation for the 
quadrupoles~\cite{flat2} . This solution is used as a
starting point for a simplex minimization algorithm that searches
the quadrupole settings to minimize the figure-of-merit $\chi ^2=
\Sigma_{13}^2+ \Sigma_{14}^2 + \Sigma_{23}^2+ \Sigma_{24}^2$,
where $\Sigma_{ij}$ is the $(ij)^{th}$ element of matrix $\Sigma$.
Upon proper tuning of the transformer, the expected normalized
flat-beam emittances, $\varepsilon_n^{\pm}$, are given
by~\cite{BND-PRE,kjk}
\begin{equation}
\varepsilon_n^{\pm}=\sqrt{({\varepsilon_n^u})^2+({\beta \gamma
\mathcal L})^2}\pm ({\beta\gamma\mathcal L}) \stackrel{\beta \gamma
\mathcal{L} \gg \varepsilon_n^u }{\longrightarrow}
\left\{\begin{array}{cc} \varepsilon_n^{+} \simeq 2 \beta \gamma \mathcal{L}  \\
 \varepsilon_n^{-} \simeq \frac{(\varepsilon_n^u)^2}{2 \beta \gamma
 \mathcal{L}} \end{array} \right. \label{eq:kjk}
,
\end{equation}
where $\varepsilon_n^u=\beta\gamma\varepsilon_u$ is the normalized
uncorrelated emittance of the  magnetized beam prior to the
transformer, $\beta= v/c$, $\gamma$ is
the Lorentz factor, ${\cal L} = \mean{L}/2p_z$, and $p_z$ is
the longitudinal momentum. Note that 
$\varepsilon_n^+ \varepsilon_n^-  =(\varepsilon_n^u)^2$.

The flat-beam emittances are measured using the slit
method~\cite{lejeune}. A movable single-slit assembly (either
vertical or horizontal), located at position X7 (see
Fig.~\ref{fig:overview}), is used to sample the beam in one
direction. The slit assembly consists of a $\sim50$~$\mu$m slit
made of a 3~mm thick tungsten block. The beamlet passing through
the slit is observed after a drift of distance $D$, at the
location X8. Given the measured horizontal beam size at X7,
$\sigma_x^{X7}$, and horizontal rms size of the beamlet at X8 when
a vertical slit is inserted at X7, $\sigma_x^{X8,h}$, the
horizontal emittance is then computed as the product
$\varepsilon_n^x = \gamma \sigma_x^{X7} \sigma_x^{X8,h}/D$.
Similarly the vertical emittance is measured as $\varepsilon_n^y =
\gamma \sigma_y^{X7} \sigma_y^{X8,v}/D$ where $\sigma_y^{X8,v}$ is
the vertical rms size of the beamlet at X8 when a horizontal slit
is inserted at X7. The beam viewer at locations X7 is an optical
transition radiation (OTR) foil, while at X8 it is a yttrium
aluminum garnet (YAG) screen. The measured rms beam size,
$\sigma_{meas}$, is affected by the resolution of the diagnostics
$\sigma_{res}$ and spurious dispersion $\eta$ introduced, e.g., by
steering dipoles required to keep the beam centered along the
beamline axis: $\sigma_{meas}=\sqrt{\sigma^2 + \sigma_{res}^2 +
(\eta \sigma_{\delta})^2}$, where $\sigma_{\delta}$ is the rms
fractional momentum spread of the beam. The measurement
method used to report emittances in the following was numerically
benchmarked~\cite{yinephD}. The resolution of the beam size
measurement system which includes the optical components and a
charged coupled device (CCD) camera was characterized for various
operating points~\cite{yinephD}. For all the quoted
measurements of transverse beam sizes, we quadratically subtract
the smallest measured resolution ($\sigma_{res}=35$~$\mu$m). The 
unavoidable contribution from spurious dispersion (discussed later) 
results in an overestimated value for the smaller flat-beam emittance. 
Hence the emittance ratio reported hereafter is underestimated. 

For the flat-beam experiment reported in this Letter, the nominal
operating parameters for the photoinjector are reported in
Table~\ref{tab:jan062005}. The rf-gun and booster-cavity settings
are kept the same during the experiment while the drive-laser spot size 
on the photocathode and the solenoid currents are adjusted for the
different sets of measurements.
\begin{table}[h!]
\begin{center}
\caption{\label{tab:jan062005} Nominal settings for the photocathode
drive laser, rf-gun and booster cavity during the flat-beam
experiment.}
\begin{tabular}{l c c}
\hline \hline parameter     &value & unit  \\
\hline
laser injection phase               &  25   & degree \\
rms laser spot size on cathode      &  0.75 $-$ 1 & mm\\
rms laser pulse duration (Gaussian) &  $\sim$3    & ps \\
bunch charge                    &  0.5  & nC \\
accelerating gradient on cathode                &  32   & MV/m     \\
axial magnetic field on cathode                &  400 $-$ 900  & Gauss     \\
booster cavity peak electric field    &  23   & MV/m     \\
\hline \hline
\end{tabular}
\end{center}
\end{table}

Given the experimental conditions, numerical simulations
are performed with the tracking program {\sc Astra}~\cite{astra}. Using the
simulation outputs of the beam properties at the entrance of the
transformer, the aforementioned simplex minimization algorithm is
used to determine the skew quadrupole settings needed to transform
the magnetized round beam into a flat beam. In the experiment, the
quadrupole settings are then empirically fine-tuned to insure the
$x$-$y$ correlation on the beam has been removed downstream of the
transformer. This is achieved by observing the beam transverse
image on the viewers downstream of the transformer: upon removal
of the angular momentum, the beam should remain flat and upright.
In Table~\ref{tab:quadIjan062005} we compare, for two cases of rms
drive-laser spot sizes ($\sigma_c$=0.76~mm and
$\sigma_c=$0.97~mm), the final quadrupole currents used in the
experiment with the initial values obtained numerically. Most of
the quadrupole currents agree with predicted values, the larger
discrepancies observed for the settings of the last quadrupole 
reflect a looser tolerance on this quadrupole setting~\cite{yinesunpac2005}.
%
%
%
\begin{table}[hbt]
\begin{center}
\caption{Comparison of the experimental skew quadrupole currents
 with the numerical predictions for different laser spot
sizes $\sigma_c$. I$_i$ is the current of the skew quadrupole
S$_i$.}\label{tab:quadIjan062005}
\begin{tabular}{c c c c c}
\hline \hline quadrupole & \multicolumn{2}{c}{$\sigma_c=0.79$~mm} &
\multicolumn{2}{c}{$\sigma_c=0.97$~mm}\\ \cline{2-5}
current&experiment & simulation & experiment & simulation\\
\hline
I$_1$(A) &-1.92&-2.03 &-1.97&-1.98 \\
I$_2$(A) &\phantom{-}2.40&\phantom{-}2.57 &\phantom{-}2.56&\phantom{-}2.58 \\
I$_3$(A) &-2.99&-4.01 &-4.55&-5.08 \\
\hline \hline
\end{tabular}
\end{center}
\end{table}

\begin{figure}[hbt]\centering
\includegraphics[height=35mm]{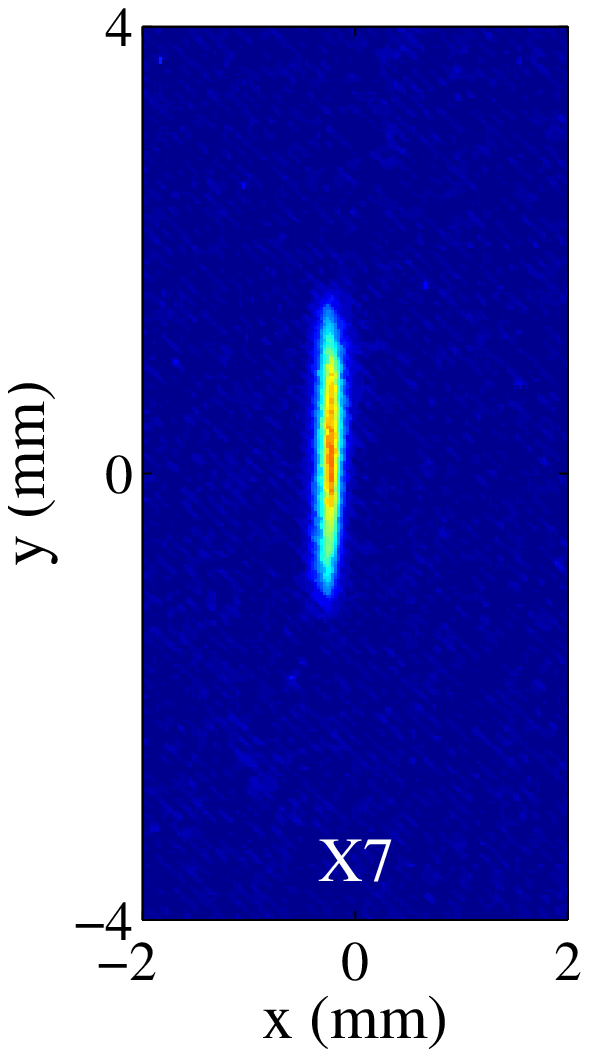}
\includegraphics[height=35mm]{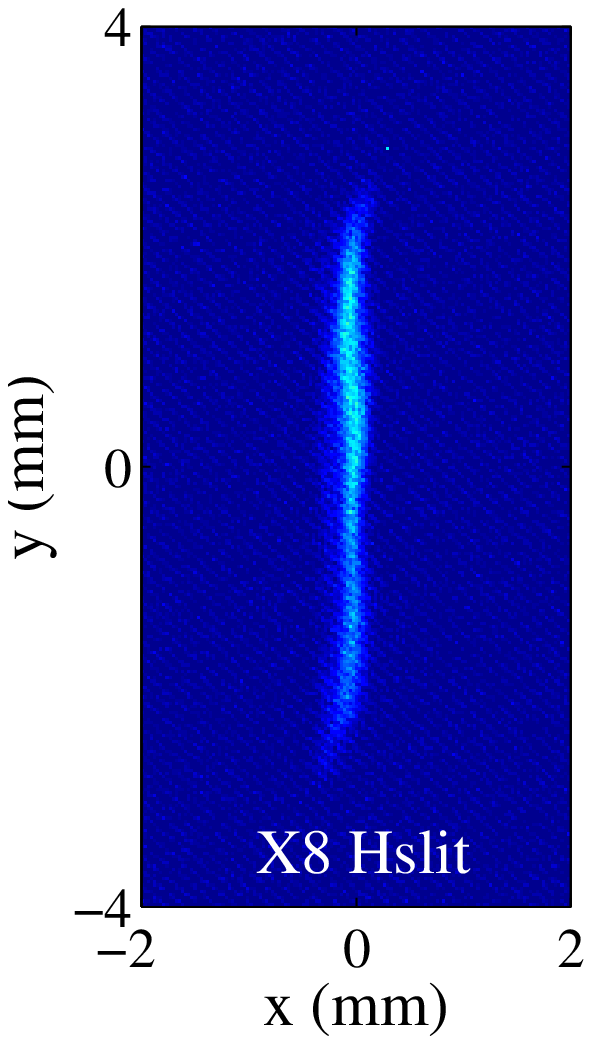}
\includegraphics[height=35mm]{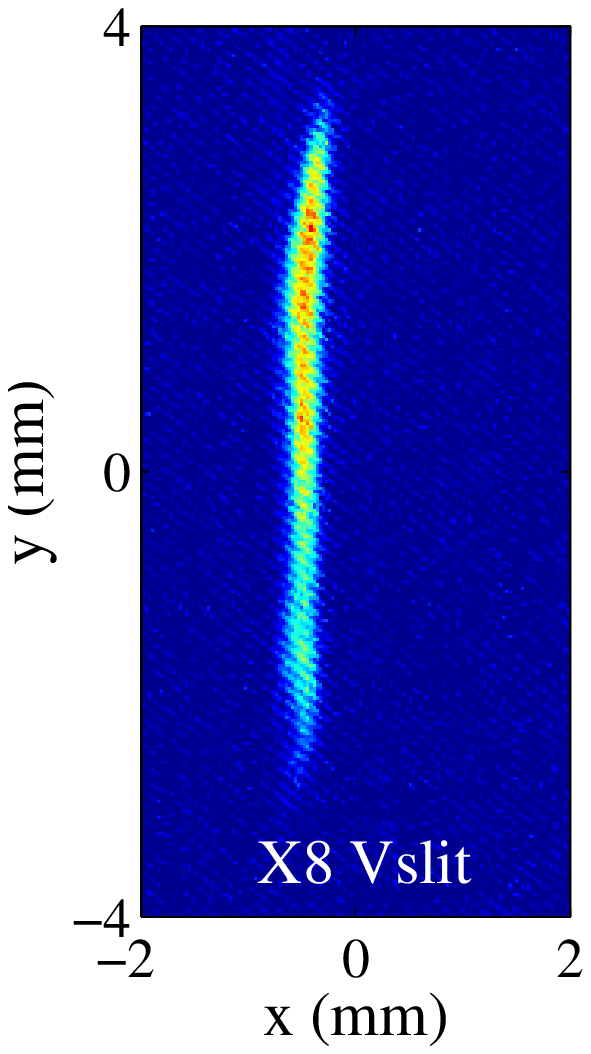}
\includegraphics[height=35mm]{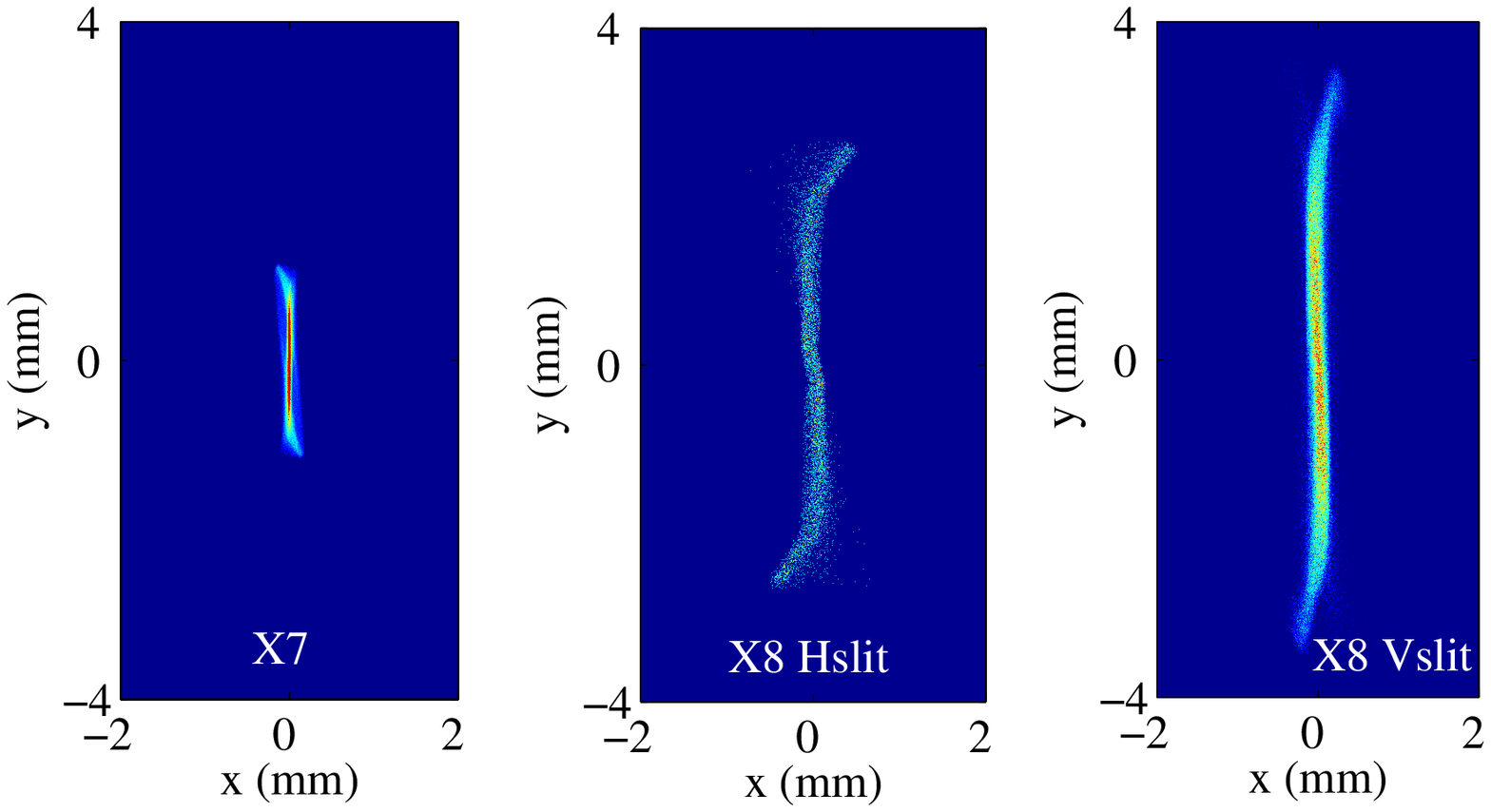}
\caption{Top three images are taken with digital cameras: beam
at X7, horizontal and vertical slit images at X8. Bottom three are
the corresponding beam profiles from \astra simulations. These
images are associated with the flat-beam presented in
Table~\ref{tab:rmstable}.} \label{fig:x7x8pics}
\end{figure}

For the transverse emittance measurements, the beam images on the
different viewers are taken for a single-bunch beam. In
Figure~\ref{fig:x7x8pics}, we present the set of experimental images,
along with their respective simulated images, needed to infer the two
transverse flat-beam emittances. Several shots of each of the
particular images are taken and analyzed to obtain the rms beam
sizes. The results are then averaged and a statistical error is
attributed to the mean. Given the uncertainty of the measurement 
method the systematic errors are estimated from error propagation. 
The rms beam sizes are estimated on 95\% of the total integrated 
image intensity. In Table~\ref{tab:rmstable}, we gather the measured and simulated
parameters for the case of $\sigma_c=0.97$~mm.
\begin{table}[htb]
\begin{center}
\caption{Measured and simulated flat-beam parameters for $\sigma_c=0.97$~mm. 
Both systematic and statistical (in brackets) errorbars are included.}\label{tab:rmstable}
\begin{tabular}{lcccc}
\hline \hline
{parameter}&
experiment & simulation &{unit} \\
\hline
$\sigma_x^{X7}$   &  0.088$\pm$0.01 ($\pm$0.01)   & 0.058 & mm\\
$\sigma_y^{X7}$   &   0.63$\pm$0.01 ($\pm$0.01)   & 0.77 &mm\\
$\sigma_x^{X8,v}$ &   0.12$\pm$0.01 ($\pm$0.01)   & 0.11 &mm\\
$\sigma_y^{X8,h}$ &   1.68$\pm$0.09 ($\pm$0.01)   & 1.50 &mm\\
$\varepsilon_n^x$ &   0.41$\pm$0.06 ($\pm$0.02)   & 0.27 &$\mu$m \\
$\varepsilon_n^y$ &   41.1$\pm$2.5  ($\pm$0.54)   & 53   &$\mu$m\\
\hline
$\varepsilon_n^y/\varepsilon_n^x$&100.2$\pm$20.2 ($\pm$5.2)& 196 & $-$\\
\hline \hline
\end{tabular}
\end{center}
\end{table}
The smaller of the flat beam emittance is $\varepsilon_n^x = 0.41\pm0.06$~$\mu$m; 
this is less than half of the expected thermal emittance due to the photoemission 
process of the cesium telluride  material. From~\cite{pitz,themit}, we infer the 
thermal emittance to be $\varepsilon_{th}=0.99\pm 0.10$~$\mu$m given 
 $\sigma_c=0.97 \pm 0.05$~mm.

%
%

\begin{table}[htb]
\begin{center}
\caption{Parameters measured from the angular-momentum-dominated
round beam and the corresponding flat beam.}\label{tab:feb252005fbcam}
\begin{tabular}{lccc}
\hline \hline parameters  & round-beam &flat-beam  &simulation\\
\hline
$\beta\gamma\mathcal{L}$ &25.6$\pm$2.6& $-$       & 26.3\\
$\varepsilon_n^u$ & 5.1$\pm$0.9   & $-$           & 3.8\\
$\varepsilon_n^+$ & 53.8$\pm$5.4\footnote{expected value
given the measured round beam parameters.}  & 41.0$\pm$2.5  & 53\\
$\varepsilon_n^-$ & 0.49$\pm$0.22$^{a}$ & 0.41$\pm$0.06 & 0.27\\
$\sqrt{\varepsilon_n^+\varepsilon_n^-}$ &5.1$\pm$0.9 & 4.1$\pm$0.8 &3.8\\
\hline \hline
\end{tabular}
\end{center}
\end{table}
\begin{figure}[h!!!!]\centering
\includegraphics[width=0.40\textwidth]{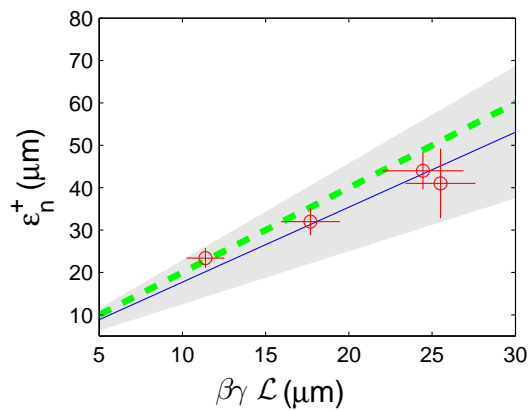}
\caption{Larger one of the flat beam emittances
($\varepsilon^+_n$) versus $\beta\gamma\cal L$. A linear
regression (solid line) of the experimental data (circle) is
compared with the theoretical dependence (dashed line). The shaded
area represents the 95\% confidence bounds associated with the
linear regression.}\label{fig:emitvsL}
\end{figure}
To gain more insight into the round-to-flat-beam transformation, 
we compare the expected flat-beam emittances, $\varepsilon^n_{\pm}$ 
in Eq.~(\ref{eq:kjk}), given the incoming magnetized beam parameters, 
with the measured flat-beam emittances downstream of the transformer. 
The uncorrelated emittance of the 
magnetized beam $\varepsilon_n^u$ is measured using the slit technique 
from the beam image at X3 and the corresponding slit images at X5. 
${\mathcal L}$ has been obtained with the two different methods
detailed in~\cite{yineprstab}. The resulting measurements for the case 
$\sigma_c=0.97$~mm are summarized in Table~\ref{tab:feb252005fbcam}: within 
the experimental errors we observed that the measured four-dimensional (4-D)
emittance $\varepsilon_{4D}\equiv\sqrt{\varepsilon_n^x \varepsilon_n^y}$ is 
conserved during the round-to-flat-beam transformation. We note a
$\sim$25\% discrepancy for the measured larger flat-beam
emittance, compared to the simulation and the value predicted from
the round beam parameters. This is probably due to imperfectly
optimized settings for the transformer. We finally report the dependence 
of $\varepsilon_n^+$ versus ${\cal L}$. The value of ${\cal L}$ was varied 
either by changing $B_0$ or $\sigma_c$. As expected $\varepsilon^+_n$ is 
linearly dependent on ${\cal L}$, and a linear regression gives 
$\varepsilon^+_n = (1.78 \pm 0.26) {\cal L}$; see Fig.~\ref{fig:emitvsL}. 
The slope is in agreement with the theoretically expected slope value of 
2 in the limit ${\cal L} \gg \beta\gamma\varepsilon_n^u$; see Eq.~(\ref{eq:kjk}).

In summary we generated and characterized a highly asymmetric beam in
a photoinjector. The lower limit for the best measured emittance 
ratio of  $\sim$100 is limited by our experimental set-up: the fact that 
the transformation occurs at low energy along with $\sigma_{\delta}\simeq 0.25$\% 
made our measurement sensitive to spurious dispersion. Simulations based 
on steering dipole settings used to correct the beam orbit indicate that the 
thereby introduced dispersion could result in an overestimation of the smaller flat-beam
emittance by a factor up to 2. Spurious dispersion accounts for most of the discrepancy 
between numerical simulations and measurements. The experiment is limited 
to low charge in order to avoid space charge to significantly impact the beam dynamics 
in  the transformer at 16~MeV. Nonetheless our  measurements support the potential 
flat-beam injector designs either for proposed light source such as LUX or envisioned 
Terahertz radiation sources based on Smith-Purcell effect. Our results also 
open a possible path for the production of flat e$^-$-beam for the ILC, where the 
main challenge is to also achieve a 4-D emittance $\varepsilon_{4D}
\sim 0.3$~$\mu$m for $Q=3.2$~nC. This value is one order of magnitude lower than what 
our photoinjector can presently produce at $Q=0.5$~nC.

We wish to acknowledge C. L. Bohn, D. Edwards, and H. Edwards, 
for encouragements, fruitful discussions and comments. We are thankful to 
J. Li and R. Tikhoplav for improving the photocathode laser.

\end{document}